\documentclass[twocolumn,showpacs,floatfix,nofootinbib,preprintnumbers]{revtex4-1}

\usepackage{amsmath, amssymb, amsfonts}
\usepackage{graphicx}
\usepackage{hyperref}
\usepackage{subfigure}
\usepackage{epsfig}
\epsfclipon

\begin{document}
 

\title{Search for doubly charged Higgs bosons\\ through  vector boson fusion  at the LHC and beyond} 

\author{G. Bambhaniya}
\email{gulab@prl.res.in}
\affiliation{Theoretical Physics Division, Physical Research Laboratory, Ahmedabad-380009, India}
\author{J. Chakrabortty}
\email{joydeep@iitk.ac.in}
\affiliation{Department of Physics, Indian Institute of Technology, Kanpur-208016, India}
\author{J. Gluza}
\email{janusz.gluza@us.edu.pl}
\author{T. Jeli\'nski} 
\email{tomasz.jelinski@us.edu.pl}
 \affiliation{Institute of Physics, University of Silesia, Uniwersytecka 4, 40-007 Katowice, Poland}
\author{R. Szafron}
\email{szafron@ualberta.ca}
\affiliation{Department of Physics, University of Alberta, Edmonton, AB T6G 2E1, Canada}

\begin{abstract}
%Boson fusion is an alternate mechanism to pin down the presence of doubly charged scalars. 
Production and decays of doubly charged Higgs bosons at the LHC  and future hadron colliders triggered by  vector boson fusion  mechanism are discussed in the context of  the Minimal Left-Right Symmetric Model. Our analysis is based on the Higgs boson mass spectrum compatible with available constraints  which include FCNC effects and vacuum stability of the scalar potential.  
Though the parity breaking scale $v_R$ is large ($\sim$ few TeV) and scalar masses which contribute to FCNC effects are even larger,  consistent Higgs boson mass spectrum still allows us to keep doubly charged scalar masses below 1 TeV which is an interesting situation for LHC and future FCC colliders. 
We have shown that the allowed Higgs bosons mass spectrum constrains the splittings  ($M_{H_{1}^{\pm \pm}}-M_{H_{1}^\pm}$), closing the possibility of $H_{1}^{\pm\pm}\to W_{1}^\pm H_{1}^\pm$ decays.
Assuming that doubly charged Higgs bosons decay predominantly into a pair of same sign charged leptons through the process $p p \rightarrow H_{1/2}^{\pm \pm} H_{1/2}^{\mp \mp} j j \rightarrow \ell^{\pm} \ell^{\pm} \ell^{\mp} \ell^{\mp} jj$, we find that for LHC operating at $\sqrt{s}=14$ TeV with an integrated luminosity at the level of $3000\,\mathrm{fb}^{-1}$ (HL-LHC), there is practically no chance to detect such particles at the reasonable significance level through this channel. However, at 33 TeV HE-LHC and (or) 100 TeV FCC-hh a wide
region opens up for exploring  the doubly charged Higgs boson mass spectrum. In
FCC-hh,   doubly charged Higgs bosons mass up to 1 TeV can be easily probed.
%
%   to have a chance of detecting even relatively light doubly charged Higgs bosons with masses up to $m_{H_{1/2}^{\pm \pm}}=500\,\mathrm{GeV}$ through the process $p p \rightarrow H_{1/2}^{\pm \pm} H_{1/2}^{\mp \mp} j j \rightarrow \ell^{\pm} \ell^{\pm} \ell^{\mp} \ell^{\mp} jj$ and with significance at the level of $3$. For much higher energies, $\sqrt{s}=33\,(100)$ TeV, it is possible to detect the considered signals at the significance level of $5$ with an  integrated luminosity $\sim1000\,(3000)\,\mathrm{fb}^{-1}$. 
\end{abstract}

%\preprint{LPN13-..., Alberta ...}

 
\maketitle

\section{Introduction}
%\section{Process}
Weak vector boson fusion (VBF)  processes were suggested quite some time ago in the context of Higgs searches \cite{Cahn:1983ip, Rainwater:1997dg, Rainwater:1998kj}. They are  characterized by the presence of two  jets with large transverse momentum $(p_T)$ in the forward region in opposite hemispheres along with other observables, like charged leptons. 
%VBF cuts require that two jets are well separated and characterized by high values of $p_T$.   
In fact, many interesting Standard Model (SM) processes, e.g. diffractive interactions, low-x QCD physics, VBF Higgs production, photo-production  are also accompanied by  production of forward particles. Interestingly  LHC has a very rich ``forward physics" program and for the necessary investigation  there are dedicated detectors like LHCf \cite{Kawade:2013bva} and(or)  FP420 \cite{Albrow:2008pn}. Due to uncertainties in jet tagging the efficiency is relatively low and, thus, the significance of these channels is rather suppressed. Nevertheless, from the discovery perspective, many Beyond Standard Models (BSM) can  also be tested using forward jets. Such related studies are also important  for  dark matter searches through mono-jet plus missing energy \cite{Fox:2011pm, ATLAS:2012zim, Khachatryan:2014rra}.

In this paper we continue \cite{Chakrabortty:2012pp,Bambhaniya:2013wza,Bambhaniya:2014cia} a dedicated analysis of the Minimal Left-Right Symmetric Model (MLRSM) \cite{Mohapatra:1974gc,Senjanovic:1975rk,Mohapatra:1986uf} aiming at exhaustive exploration of interesting BSM signals at present and future hadron colliders\footnote{The main features of this model are equal $SU(2)$ left and right gauge couplings, $g_L=g_R$, and a scalar potential which contains a bidoublet and two triplet scalar multiplets, considered for the first time in \cite{Mohapatra:1980yp}, see also \cite{Gunion:1989in,Duka:1999uc}.}. 
For many reasons the  parity breaking scale $v_R$ of the right $SU(2)$ 
group in MLRSM must be already around ${\mathcal{O}}(10)$ TeV  \cite{Chakrabortty:2012pp,Bambhaniya:2013wza,Bambhaniya:2014cia,Zhang:2007da,Melfo:2011nx}. However, as discussed recently in \cite{Bambhaniya:2013wza,Bambhaniya:2014cia},  in such models  charged Higgs bosons can have masses at the much lower level of a few hundred GeV, and that scenario is still consistent with experimental data. In this case, it is imperative to cover 
all possible scenarios, and their   potential effects at the LHC should be analyzed carefully. Interestingly enough, a recent CMS study \cite{Khachatryan:2014dka} 
can be interpreted as favoring right handed currents. 

In our previous analyses we have worked with the scalar mass spectra  which are compatible with the unitarity of the potential parameters, 
the large parity breaking scale $v_R$ and the severe bounds on neutral scalar masses ($M_{H_1^0},M_{A_1^0}$) derived from Flavor Changing Neutral Currents (FCNC). In this work we have further implemented another necessary condition:  vacuum stability of the scalar potential. It appears that even after taking into account all these constraints, the consistent scalar mass spectra can accommodate doubly charged Higgs boson masses in a region which can be explored by the LHC. 

In the past, we have focused on searches for  multi-lepton signals associated with any number of jets; i.e., there was no jet veto.  Here,  the  analysis of possible VBF-type signals with four leptons and two jets using suitable VBF cuts is presented. 

We have used  our version of the Left-Right symmetric model implemented in FeynRules (v2.0.31) \cite{Christensen:2008py,Degrande:2011ua}. The general signal and background analyses for multi-lepton and tagged forward jets are performed using   ALPGEN (v2.14) \cite{Mangano:2002ea}, Madgraph (v2.2.2) \cite{Alwall:2011uj} and PYTHIA (v6.421) \cite{Sjostrand:2006za}.  

%To make our analysis more realistic we have used the correlations among the model parameters and generate the consistent spectrum compatible with low energy data and safe from FCNCs \cite{Czakon:1999ue,Czakon:1999ga,Duka:1999uc,Chakrabortty:2012pp}.

\section{Possible processes which identify doubly charged Higgs through VBF in MLRSM}

\begin{figure*}[htp]
%\begin{center}
\subfigure{\includegraphics[width=0.35\linewidth]{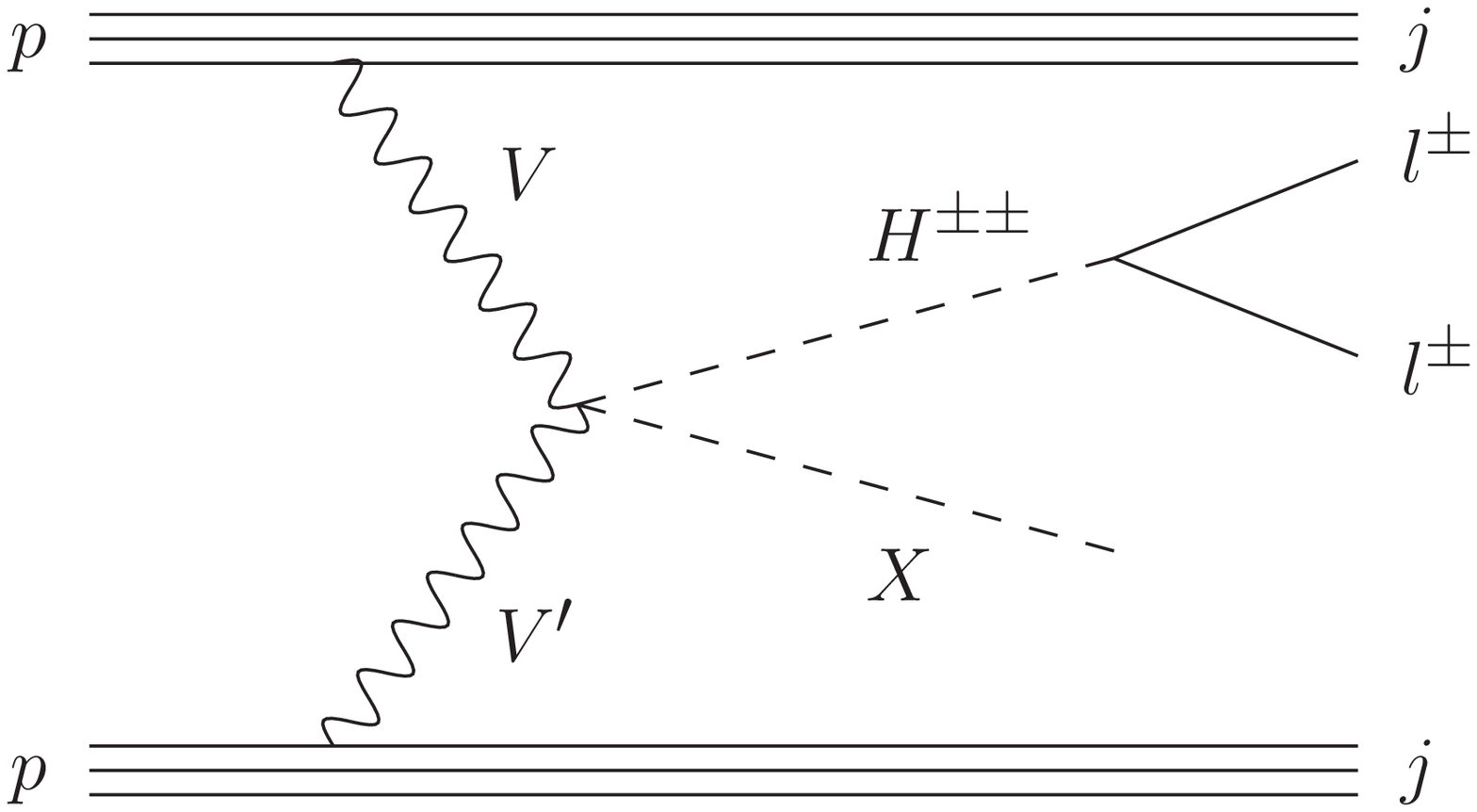}}\quad 
\subfigure{\includegraphics[width=0.35\linewidth]{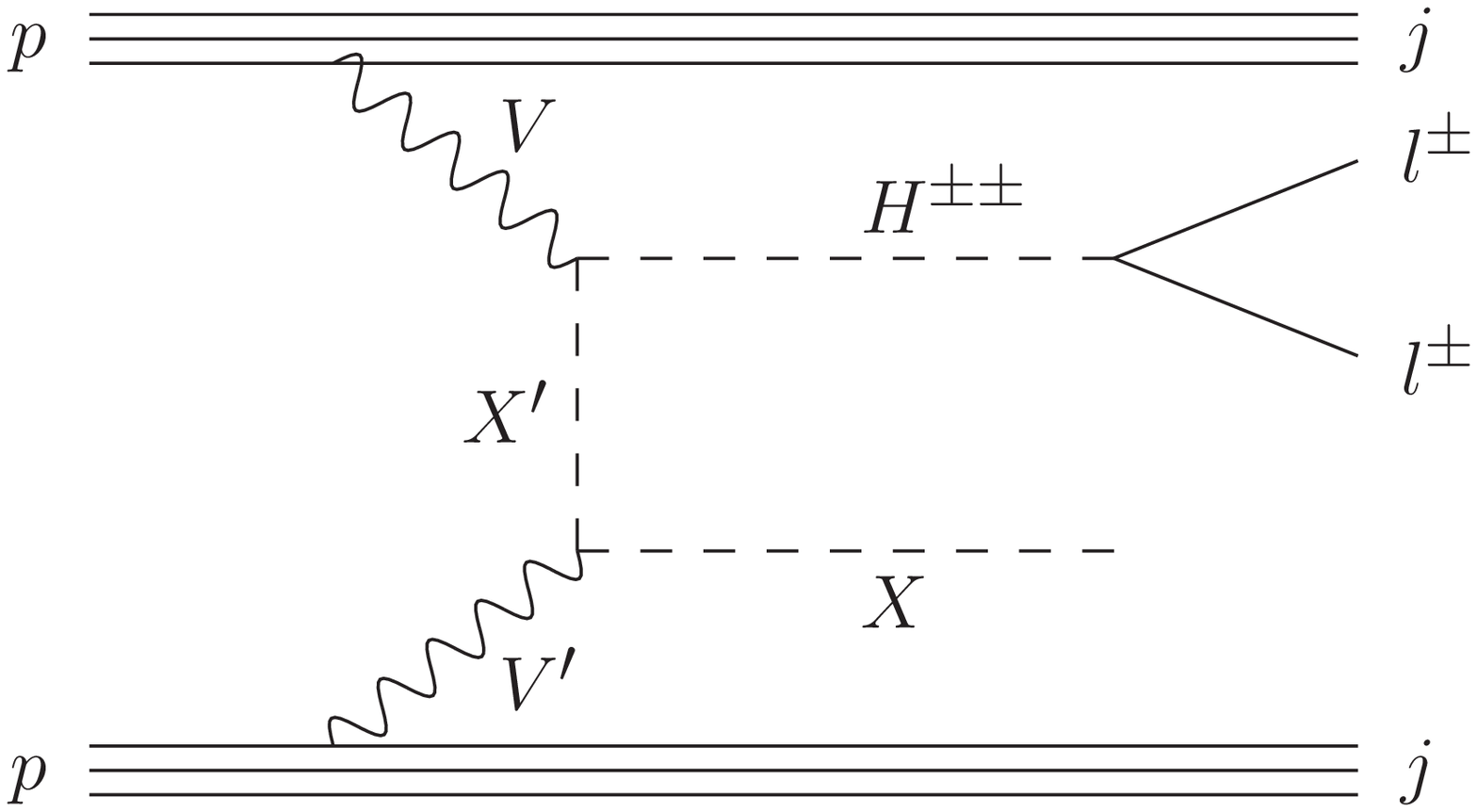}}\\
\subfigure{\includegraphics[width=0.35\linewidth]{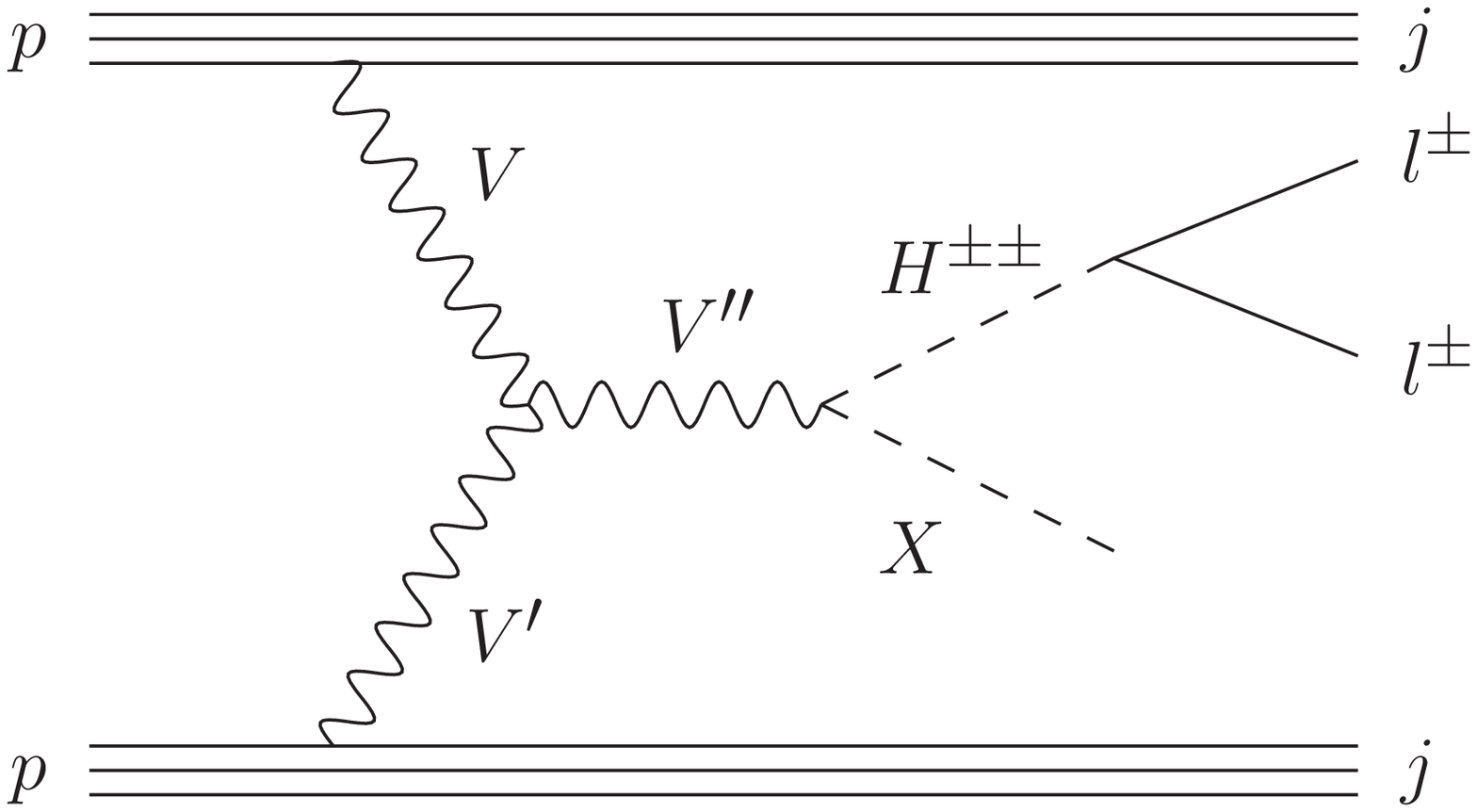}}\quad 
\subfigure{\includegraphics[width=0.32\linewidth]{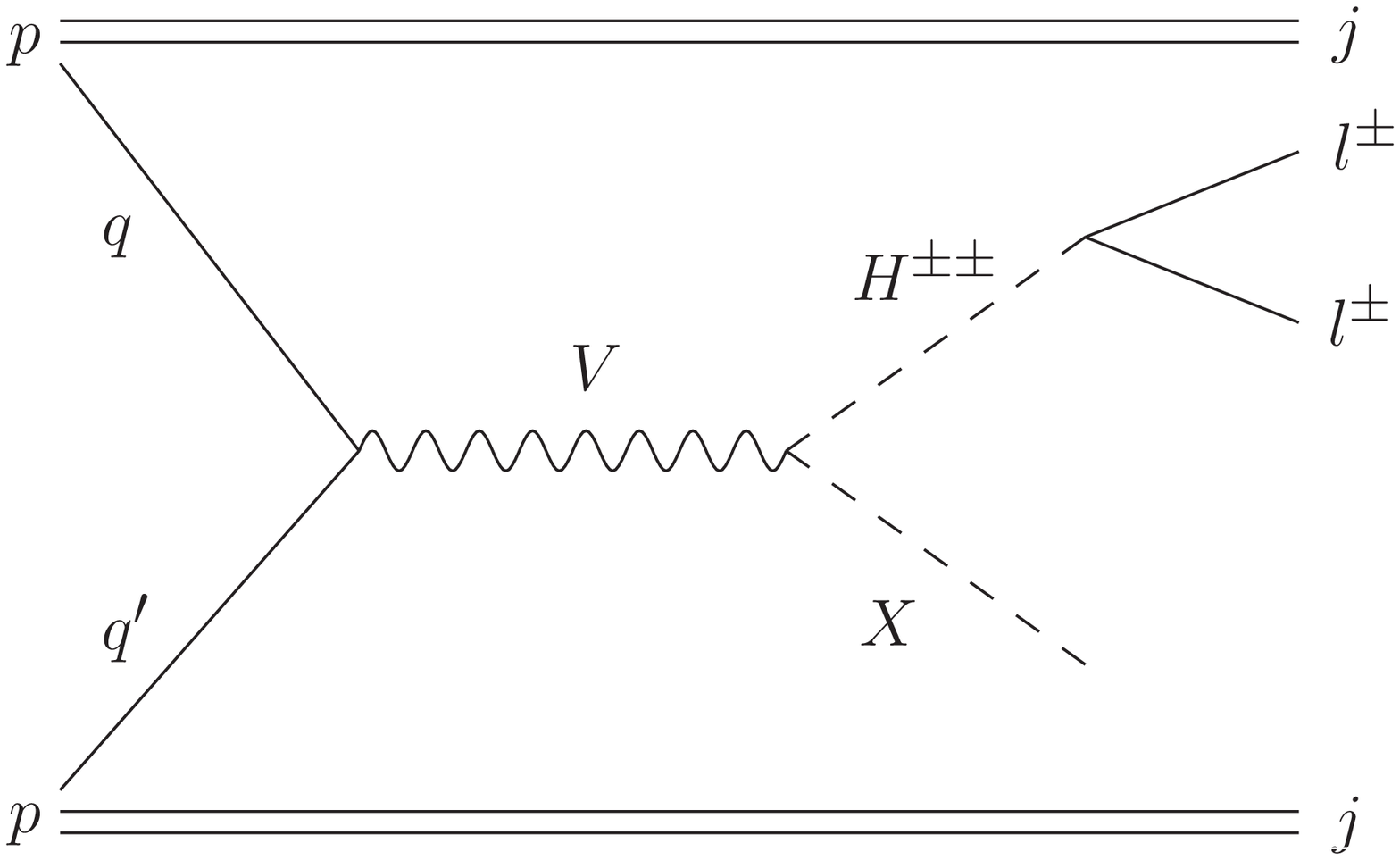}}
%
%
%
%\includegraphics[width=0.76\linewidth]{ppHpmpmX4new.eps}
%%\label{fig:VBFHpmpmXa}
%\\
%(a)\\
%\vspace{5mm}
%\includegraphics[width=0.76\linewidth]{ppHpmpmXtwo3new.eps}
%%\label{fig:VBFHpmpmXb}
%\\
%(b)\\
%\vspace{5mm}
%\includegraphics[width=0.76\linewidth]{ppHpmpmX3bosnew.eps}
%%\label{fig:VBFHpmpmXc}
%\\
%(c)\\
%\vspace{5mm}
%\includegraphics[width=0.76\linewidth]{qqpHpmpmXnew.eps}
%%\label{fig:VBFHpmpmXd}
%\\
%(d)
%\end{center}
\caption{\small Basic processes which lead to $H^{\pm\pm}$ pair production. In the first three diagrams $H^{\pm\pm}$  is produced through fusion of two vector bosons $V$ and $V'$. Each of them can be $W^{\pm}$, $Z^0$ or $\gamma$. The second  product of the fusion, scalar $X$, is $H^{\pm\pm}$, $H^{\pm}$ or $H^0$, depending on the configuration of colliding vector bosons. Analogously, scalar $X'$ and vector boson $V''$ can be identified once $V$ and $V'$ are specified.  In the last diagram $H^{\pm\pm}$ is produced through collision of two quarks $q$ and $q'$ in the Drell-Yan process. The second  product of the decay, scalar $X$, can be identified as $H^{\pm\pm}$, $H^{\pm}$ or $H^0$ once $V$ is specified.}
\label{fig:VBFHpmpmX}
\end{figure*}

There are many interesting channels in which doubly charged Higgs particles can be produced in MLRSM. 
In the hadron collider, productions of doubly charged Higgs particles crucially depend on their couplings with vector bosons. These charged scalars ($H^{\pm\pm}$) are produced either through neutral and charged currents or fusion processes. 
Representative classes of diagrams which contribute to $H^{\pm \pm}$ productions associated with two jets are given in  Fig.~\ref{fig:VBFHpmpmX}.

If $X=H^{\pm\pm}$ in Fig.~\ref{fig:VBFHpmpmX}, then doubly charged Higgs particles are produced in pairs. Assuming further that $H^{\pm\pm}$ decays predominantly into leptons, a signal of four leptons associated with two forward jets in the final state is foreseen, $p p \rightarrow H_{1/2}^{\pm \pm} H_{1/2}^{\mp \mp} j j \rightarrow \ell^{\pm} \ell^{\pm} \ell^{\mp} \ell^{\mp} jj$. In the Drell-Yan case  also [see the diagram (d) in Fig.~\ref{fig:VBFHpmpmX}], if $X = H^{\pm\pm}$,  a four leptons plus two jets signal is possible, though its contribution is suppressed once the VBF cuts are activated.  

We should also mention that  vector boson fusion diagrams  interfere substantially with Bremsstrahlung-like (or Drell-Yan) processes \cite{Massironi:2014fma}.

Here we focus  on the pair production of doubly charged scalars associated with two forward jets. As mentioned already, this signature can be promising  since LHC has dedicated search channels for tagged forward jets. VBF processes with doubly charged Higgs bosons have been considered lately in \cite{Dutta:2014dba} with the main focus on three lepton signals with missing energy and in \cite{Englert:2013wga} where doubly charged Higgs bosons decay into same-sign $W$ bosons. In  \cite{Dutta:2014dba} there is also an interesting discussion  on scalar corrections to the  self-energy of $W_L^\pm$ and the $\Delta \rho_{EW}$ parameter. It has been argued  that there exist severe constraints on the charged scalar mass splitting. However, in our opinion
conclusions based on partial results and a single class of (scalar) diagrams can be deceptive; thus, we need to  consider complete calculations including renormalization. We recall a series of papers on the 1-loop corrections to the muon decay in MLRSM, starting with qualitative results \cite{Czakon:1999ue,Czakon:1999ha} and finishing with quantitative analysis  \cite{Czakon:2002wm}. 
%More general analysis concerning models which have $\Delta \rho_{EW} \neq 1$ at the tree level are given in \cite{Czakon:1999ha}, see also \cite{Chankowski:2006hs,Chankowski:2006jk,Chen:2008jg,Kanemura:2012rs}.
The upshot of all these analyses, important for our present discussion, is that  there is a strong fine-tuning between contributions to $\Delta \rho_{EW}$ from different classes of non-standard particles: Higgs and additional gauge {\it bosons} and heavy neutrinos 
({\it fermions}). By their nature, cancellations among bosonic and fermionic types of diagrams are present, and a change of mass spectrum of the Higgs bosons can be compensated for by different choices of $v_R$ scale (gauge bosons) and masses of heavy neutrinos. These analyses in the context of the LHC  have been considered in detail in \cite{Chakrabortty:2012pp}.
 
%In our analysis we allow for larger splittings  
 
\section{Constraints on $\alpha_3$ and $\delta\rho$: adding vacuum stability condition \label{sec:vac}}

\begin{figure}[h!]
%\begin{center}
%\includegraphics[width=5cm, angle=-90]{PLOTS/3000L/ssdl_lepton_distribution.eps}
%\includegraphics[width=5cm , angle=-90]{PLOTS/1000L/ssdl_lepton_distribution.eps} \vskip 0.1cm
\includegraphics[height=8.5cm, width=7cm , angle=-90]{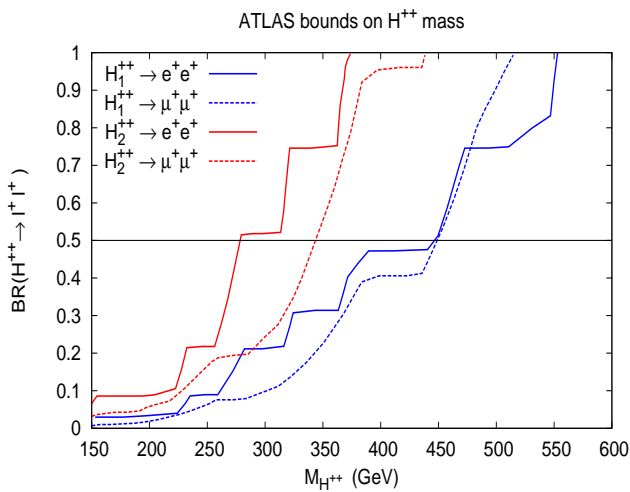}
\caption{\small Exclusion limits on the masses of doubly charged scalars from the ATLAS analysis, depending on their leptonic branching ratios. The lepton flavor violating modes are not shown here, as they are not concerned with the purpose of our analysis. This plot is based on Fig.~5 in \cite{ATLAS:2014kca}.}
\label{fig:ATLAS_excl}
%\end{center}
\end{figure}

Analysis of the LHC data provides lower limits on the doubly charged Higgs mass \cite{ATLAS:2014kca} depending on their leptonic decay branching fractions. In the scenario where ${\mathrm{BR}}(H^{++}\to e^{+}e^{+})={\mathrm{BR}}(H^{++}\to\mu^+\mu^+)\approx0.5$, that limit is $M_{\mathrm{LHC}}= M_{H^{\pm\pm}}\approx 450\,\mathrm{GeV}$; see Fig.~\ref{fig:ATLAS_excl} for details. 
 
Limits on the MLRSM potential parameters have been discussed lately in \cite{Bambhaniya:2014cia}. 
Similar to the earlier case,  we focus  on the  $\alpha_3$ and $\delta\rho=\rho_3-2\rho_1$ parameters, which are important for the scalar mass spectrum 
(all notations are as in \cite{Bambhaniya:2013wza,Bambhaniya:2014cia}).
% Let us discuss what are the constraints on the allowed values of the scalar potential parameters $\alpha_3$ and $\delta\rho=\rho_3-2\rho_1$ for the fixed value of $v_R$.  
First, to suppress FCNC effects generated by $H_1^0$ and $A_1^0$, we assume\footnote{To our knowledge, their effects have been discussed for the first time in the context of Left-Right models in \cite{Ecker:1983uh}, see also \cite{Mohapatra:1983ae, Pospelov:1996fq, Zhang:2007da, Maiezza:2010ic, Chakrabortty:2012pp} and recently \cite{Bertolini:2014sua}. In general, their masses need to be at least of the order of 10 TeV, though some alternatives also have been  considered in \cite{Guadagnoli:2010sd}.} that their mass is bigger than $M_{\mathrm{FCNC}}=10\,\mathrm{TeV}$. Because $M_{H_1^0,A_1^0}^2=\alpha_3v_R^2/2$, this results in the following lower limit on $\alpha_3$:
\begin{equation}\label{alpha-FCNC}
\alpha_3\geq\frac{2M_{\mathrm{FCNC}}^2}{v_R^2}.
\end{equation}
Taking into account that $M_{H_1^{\pm\pm}}^2=(\delta\rho~ v_R^2+\alpha_3 ~\kappa^2)/2$, one gets
\begin{equation}\label{alpha-LHC}
\alpha_3\geq\frac{1}{\kappa^2}(2M_{\mathrm{LHC}}^2-\delta\rho ~v_R^2),
\end{equation}
where $\kappa=246\,\mathrm{GeV}$ is the electroweak symmetry breaking scale. The third constraint originates from the {\it necessary condition} of the boundedness of the potential \cite{Chakrabortty:2013mha}:
\begin{equation}\label{alpha-JD} 
\alpha_3\leq\sqrt{8\lambda_1(4\pi-\delta\rho)}.
\end{equation}
The value of  $\lambda_1$ is fixed by the lightest neutral Higgs boson mass as $M_{H_0^0}^2=2\lambda_1 \kappa^2$. 
%
%

%
%As one can see these bounds suggested in  Eqs.~(\ref{alpha-FCNC}), (\ref{alpha-LHC}) and (\ref{alpha-JD}) %restrict a wide range of parameter space $(\alpha_3,\delta\rho)$ and rather allow a modest set of values.

\begin{figure*}[htp]
%\begin{center}
%\includegraphics[height=7cm, width=9cm , angle=0]{mass-split-2.eps} 
\subfigure{\includegraphics[scale=0.75]{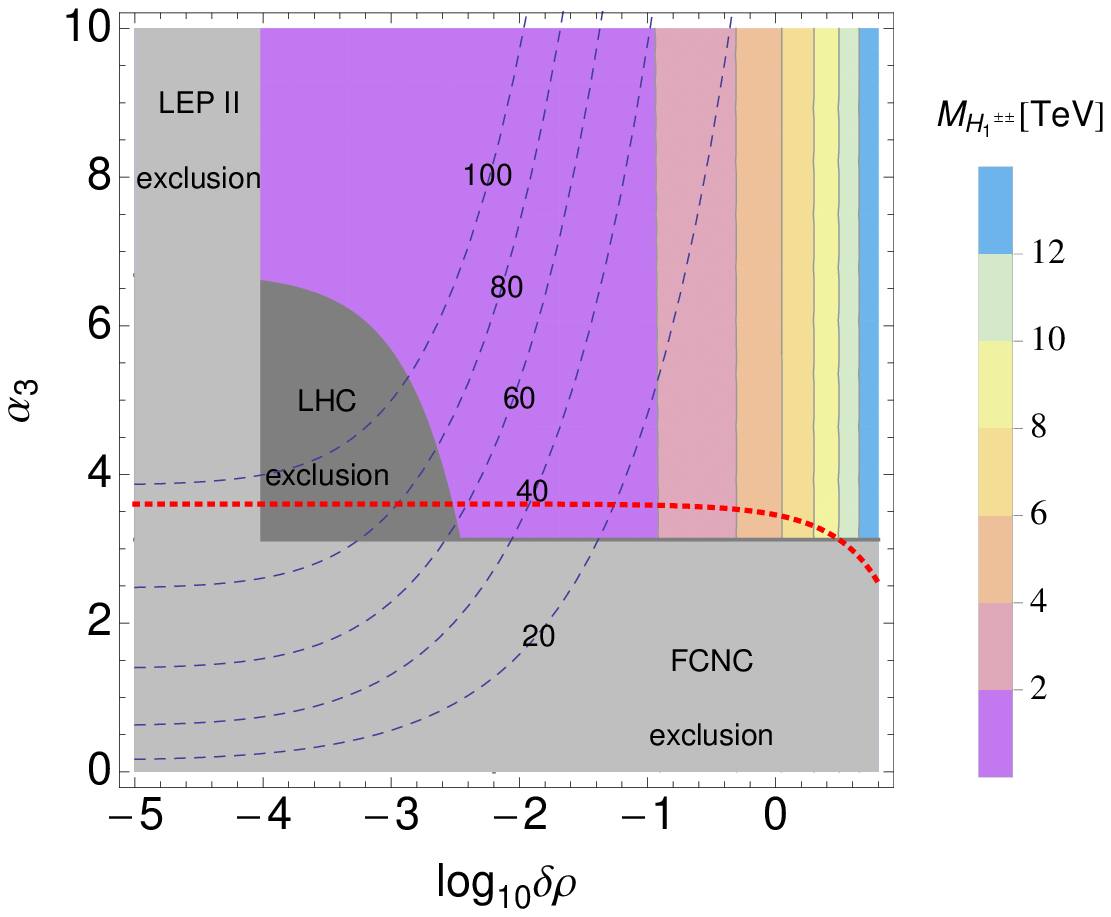}}\quad 
\subfigure{\includegraphics[scale=0.75]{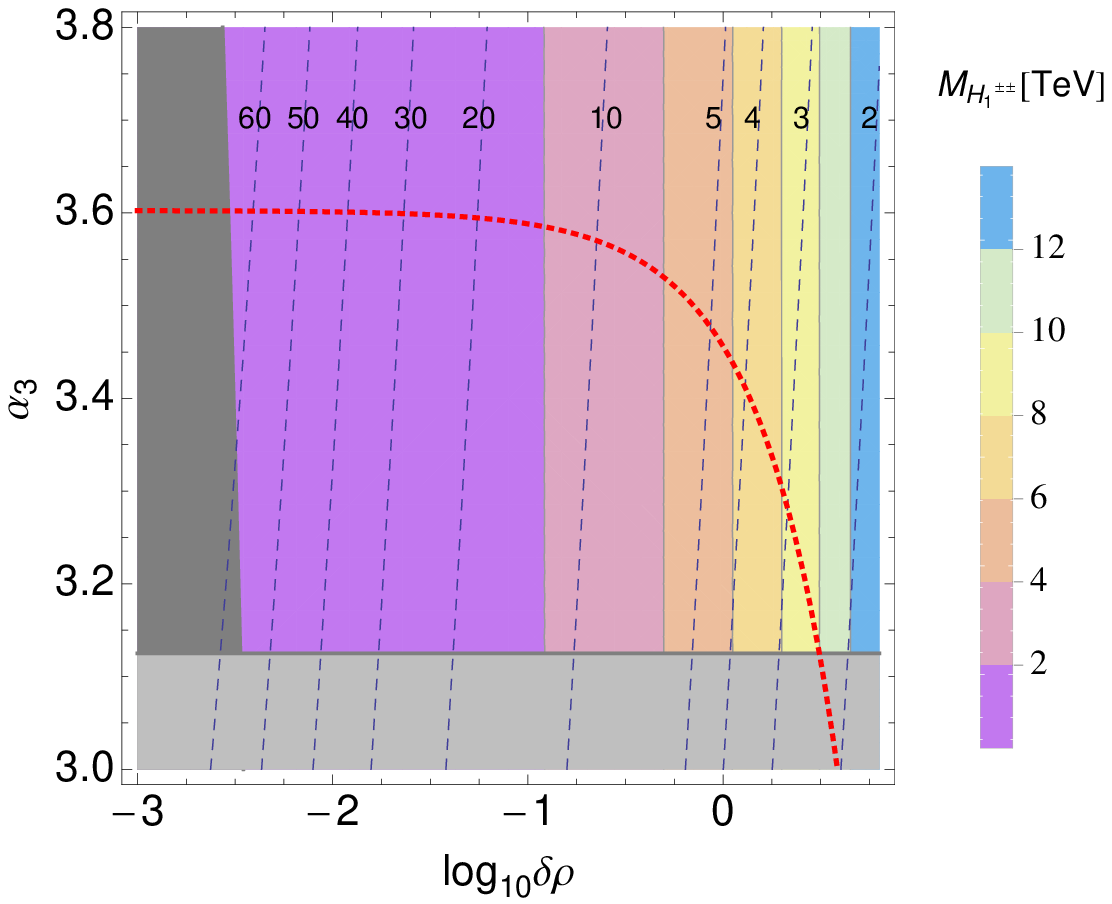}}
\caption{\small 
(\textit{left panel}) Dependence of the $H_1^{\pm\pm}$ mass (in $\textrm{TeV}$) on $\delta\rho$ and $\alpha_3$ for $v_R = 8\,\textrm{TeV}$. The parameter space $(\delta\rho,\alpha_3)$ is divided into coloured regions where mass of $H_1^{\pm\pm}$ is characterized according to the attached  legend. Shaded regions are excluded due to FCNC, LHC and LEP constraints - see Eqs.~(\ref{alpha-FCNC}) and (\ref{alpha-LHC}) and 
Refs.~\cite{Datta:1999nc, Bambhaniya:2014cia} respectively. The parameter space above the red-dotted line is disfavoured due to the unboundedness of the scalar potential - see Eq.~(\ref{alpha-JD}). Blue, dashed lines represent sets of points $(\delta\rho, \alpha_3)$ for which mass splitting $(M_{H_1^{\pm\pm}}-M_{H_1^{\pm}})$ is $100$, $80$, $60$, $40$ and $20\,\textrm{GeV}$ respectively. (\textit{right panel}) 
Detailed view of the allowed part of parameter space %shown together 
with refined mass splitting lines. 
}
\label{fig:split}
%\end{center}
\end{figure*}

Now it is interesting and important to ask what is the maximum allowed mass splitting $\Delta M=M_{H_1^{\pm\pm}}-M_{H_1^{\pm}}$ that will be consistent with the bounds on $(\alpha_3,\delta\rho)$ derived above. Such queries cannot be unnoticed  from  a phenomenological perspective because only for $\Delta M>M_{W_1}$ can the  doubly charged Higgs have the following decay: $H_1^{\pm\pm}\to H_1^{\pm}W_1^{\pm}$. This has a massive impact on the decay branching ratios of $H^{\pm \pm}$. It is straightforward to check that the biggest $\Delta M$ is reached for $\delta\rho$, saturating both inequalities in Eqs.~(\ref{alpha-LHC}) and (\ref{alpha-JD}) which imply 
\begin{equation}
\frac{1}{\kappa^2}(2M_{\mathrm{LHC}}^2-\delta\rho v_R^2)=\sqrt{8\lambda_1(4\pi-\delta\rho)}.
\end{equation}
The physical solution to this equation and the corresponding maximal value of $\Delta M$ is
 
\begin{eqnarray}
\delta\rho&=& \frac{2(M_{\mathrm{LHC}}^2-\sqrt{8\pi\lambda_1}\kappa^2)}{v_R^2}(1+\ldots),\nonumber\\ 
\Delta M&=&\Delta M_{\infty}(1+\ldots),
\end{eqnarray}
where `$\ldots$' stands for corrections of the order of $\mathcal{O}\left(M_{\mathrm{LHC}}^2/v_R^2,\kappa^2/v_R^2\right)$. One can check that $\Delta M$ depends on $v_R$ very weakly and is nearly equal to the asymptotic value $\Delta M_{\infty}=\lim_{v_R\to\infty}\Delta M=M_{\mathrm{LHC}}-\sqrt{M_{\mathrm{LHC}}^2-\sqrt{2\pi\lambda_1}\kappa^2}\approx65.3\,\mathrm{GeV}$, for $M_{\mathrm{LHC}}=450\,\mathrm{GeV}$. As $\partial_{v_R}\Delta M>0$, this implies that on-shell  decay $H_1^{\pm\pm}\to H_1^{\pm}W_1^{\pm}$ is kinematically forbidden regardless of the scale $v_R$. Interestingly, we came to the same conclusion as in \cite{Dutta:2014dba}, but based on different kinds of arguments. There is another consequence of the requirement that the scalar potential is bounded from below. Namely, one can show that, using Eqs.~(\ref{alpha-FCNC}) and (\ref{alpha-JD}), in the allowed parameter space 
%parameter space restricted by (\ref{alpha-FCNC})-(\ref{alpha-JD}) 
there is an upper limit on the $H_1^{\pm\pm}$ mass:
\begin{equation}
M_{H_1^{\pm\pm}}\leq \frac{1}{2}\sqrt{8\pi v_R^2-\frac{M_{\mathrm{FCNC}}^4}{\lambda_1 v_R^2}}(1+\ldots)\approx9.98\,\mathrm{TeV},
\end{equation}
where `$\ldots$' stands for the corrections of order of $\mathcal{O}(\kappa^2/v_R^2)$. The maximal value of $M_{H_1^{\pm\pm}}$ is reached for $\delta\rho$ satisfying $\sqrt{8\lambda_1(4\pi-\delta\rho)}=2M_{\mathrm{FCNC}}^2/v_R^2$ and $\alpha_3=2M_{\mathrm{FCNC}}^2/v_R^2$, which correspond to the intersection point of lines restricting  regions defined by Eqs.~(\ref{alpha-FCNC}) and (\ref{alpha-JD}). The situation is summarized in Fig.~\ref{fig:split}. 
%
%{\bf This is not very clear to me still 100 GeV separation is allowed? And this should be reflected in the fig and if this is correct then we are opposing the choice of parameters of ref. \cite{Dutta:2014dba}.}
%
Naturally, the minimal value of the $H_1^{\pm\pm}$ mass in the discussed setup is $M_{\mathrm{LHC}}$. For the sake of  completeness, let us note that if there are no  experimental limits on $M_{H_1^{\pm\pm}}$ and $H_3^0$, then the lowest possible mass of $H_1^{\pm\pm}$ consistent with the vacuum stability bound, Eq.~(\ref{alpha-JD}), would be $\sqrt{2\sqrt{\pi}M_{H_0^0}v}\approx330\,\mathrm{GeV}$, which corresponds to $\delta\rho\to0$ and $\alpha_3\to\sqrt{32\pi\lambda_1}$. On the other hand, the MLRSM does not provide any relevant constraints on the $H_2^{\pm\pm}$ mass\footnote{The only constraint which could arise is $M_{H_2^{\pm\pm}}< 2\sqrt{6\pi} v_R\approx40\,\mathrm{TeV}$ for $v_R=8\,\mathrm{TeV}$. It comes from the assumption that scalar potential parameter $\rho_2$ is in the perturbative regime $\rho_2<4\pi$.}.  We would like to mention that when $\rho_2$ satisfies
\begin{equation}
\rho_2<\frac{1}{4}\mathrm{min}(\alpha_3,\delta\rho)+\frac{1}{2}\frac{M_{W_{1,2}}^2}{v_R^2}-\frac{1}{8}\alpha_3\frac{\kappa^2}{v_R^2},
\end{equation}
then a similar type of decay of $H_2^{\pm \pm}$ is kinematically forbidden as 
%$W_2^{\pm}$ and 
$H_2^{\pm\pm}$ is too light to decay into $H_2^{\pm}$ and $W_{1,2}^{\pm}$, respectively; see the benchmarks in the next section.

\section{Predictions  for  $p p \rightarrow H_{1/2}^{\pm \pm} H_{1/2}^{\mp \mp} j j \rightarrow \ell^{\pm} \ell^{\pm} \ell^{\mp} \ell^{\mp} jj$ in MLRSM}\label{sec:simulation_criteria}

Before we discuss our simulated results, selection criteria should be defined, which are crucial for extracting proper signals and reducing the SM background. 
For selecting leptons, we  use the same criteria as defined in previous papers
\cite{gb-jd-sg-pk,Bambhaniya:2013wza}, which are read as follows:
\begin{itemize}
\item  Lepton identification criteria: pseudo-rapidity $|\eta_{\ell}| < 2.5$ and ${p_T}_{\ell}>10\,\mathrm{GeV}$;
\item   Detector efficiency for charged leptons: 
\begin{itemize}
\item   electron (either $e^\pm$): $0.7\,(70\%)$;
\item    muon (either $\mu^\pm$): $0.9\,(90\%)$;
\end{itemize}
\item Smearing of muon $p_T$ and  electron energy are implemented in PYTHIA;
\item Lepton-lepton separation:  $\Delta R_{ll} \ge 0.2$;
\item Lepton-photon separation: $\Delta R_{l\gamma} \ge 0.2$ where all the photons have ${p_T}_\gamma > 10\,\mathrm{GeV}$;

\item We have implemented a $Z$-veto to suppress the SM background, and  this has a larger impact while reducing the background for the four-lepton without missing energy. This veto reads as follows: The same flavored but opposite sign lepton pair invariant mass $m_{\ell_1\ell_2}$ must be sufficiently away from the SM $Z$-boson mass, say,  $|m_{\ell_1\ell_2} - M_{Z_1}| \geq 6 \Gamma_{Z_1} \sim 15$ GeV;
\item Lepton-jet separation: The separation of a lepton with all nearby jets must satisfy $\Delta R_{lj} \ge 0.4$. If this is not satisfied, then  that lepton is not counted as a lepton. For completeness, we must mention that 
jets are constructed from hadrons using PYCELL within PYTHIA;
\item Hadronic activity cut: this cut is applied to consider only those leptons that have much less hadronic activity around them.  
Each lepton should have hadronic activity which is accounted as $\frac{\sum p_{T_{hadron}}}{p_{T_l}} \le 0.2$ within the cone of radius $0.2$ around the lepton;
\item Hard $p_T$ cuts for four lepton events: ${p_T}_{l_1}>30\,\mathrm{GeV}$, ${p_T}_{l_2}>30\,\mathrm{GeV}$, ${p_T}_{l_3}>20\,\mathrm{GeV}$, ${p_T}_{l_4}>20\,\mathrm{GeV}$. 
\end{itemize}
\begin{table}[h]
\begin{center}
\begin{tabular}{|c|c|c|c|c|}
\hline 
%\hline
Cuts    & ${p_T}_{j_1},{p_T}_{j_2}$  & $|\eta_{j_1} - \eta_{j_2}|  $  &  $m_{j_1j_2}$ & $\eta_{j_1}*\eta_{j_2} $\\
%&    &   &  &    \\
\hline
VBF  & $ \geq 50$ & $> 4$ & 500 & $< 0$   \\
\hline
\end{tabular}
\caption{Selection criteria for the forward jets. The two highest $p_T$ jets ${p_T}_{j_1},{p_T}_{j_2}$ are chosen as the VBF forward jets.}
\label{table:forward_jet_selection}     
\end{center}
\end{table}
The Parton Distribution Function (PDF) for protons is defined by CTEQ6L1 \cite{Pumplin:2002vw}.
After satisfying the above selection criteria, additional  cuts are applied to identify the forward jets.  The detail of these VBF cuts are depicted in Tab.~\ref{table:forward_jet_selection}. 
\begin{figure}[h!]
%\begin{center}
%\includegraphics[width=5cm, angle=-90]{PLOTS/3000L/ssdl_lepton_distribution.eps}
%\includegraphics[width=5cm , angle=-90]{PLOTS/1000L/ssdl_lepton_distribution.eps} \vskip 0.1cm
\includegraphics[height=8.6cm, width=6.5cm, angle=-90]{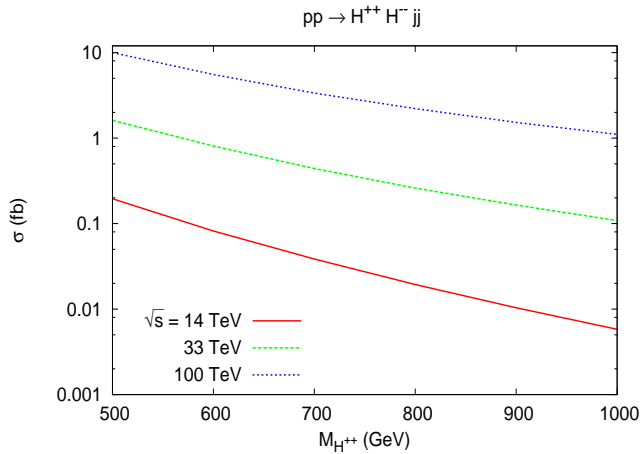} 
\caption{Dependence of cross sections ($\sigma$) with the masses of doubly charged scalars for the process $pp\to H^{++} H^{--} jj$ for different centre of mass energies: 14 TeV (red-solid), 33 TeV (green-dashed), and 100 TeV (blue-dotted), respectively.}
\label{fig:inv_mass_SSDL_OSDL_3000}
%\end{center}
\end{figure}

Considering the constraints on potential parameters discussed in section~\ref{sec:vac}, in Fig.~\ref{fig:inv_mass_SSDL_OSDL_3000} the results 
are presented for the doubly charged Higgs production process with two jets as a function of their mass.  
While computing the MLRSM mass spectrum, we have set $v_R=8\,\mathrm{TeV}$ (which leads to  $M_{W_2}=3.76\,\mathrm{TeV}$).
The analyses are performed for the LHC with $14\,\mathrm{TeV}$ collision energy considering the high luminosity HL-LHC option \cite{Barletta:2013ooa} as well as for future scenarios such as HE-LHC with center of mass energy $33\,\mathrm{TeV}$ \cite{Assmann:1284326,Barletta:2013ooa}
%AguilarSaavedra:2001rg, 
%CMS-TDR, 
%Djouadi:2007ik, Aad:2009wy
or the 100 TeV FCC-hh facility \cite{fcc,fccprep,fccbrief,fccweek2015}.
%\cite{Andeen:2013zca, Apanasevich:2013cta, 2013arXiv1309.7452D, Stolarski:2013msa, Yu:2013wta, Zhou:2013raa, Cohen:2013zla, Cohen:2013xda, Barr:2014sga, Bramante:2014tba, 100TeVpp}.
The cross section for this process has been computed with a large $p_{T_j} $ and VBF cuts as defined in Tab.\ref{table:forward_jet_selection} 

As an example of a representative Higgs mass spectrum (benchmark) used in calculations, we assume degenerate doubly charged Higgs masses $M_{H_1^{\pm \pm}}=M_{H_2^{\pm \pm}}=500 \;[1000]\,\mathrm{GeV}$ where masses of remaining scalar particles compatible with results of section~\ref{sec:vac} can be chosen as (in $\mathrm{GeV}$):
\begin{eqnarray}
M_{H^0_0} &=& 125\; [125],\label{B2MH00},\; M_{H^0_1}=  10431\; [10431],\\ 
M_{H^0_2} &=& 27011\; [27011], \;
M_{H^0_3} = 384\; [947], \\
M_{A^0_1} &=& 10437\; [10437],\; 
M_{A^0_2} = 384\; [947],\\
M_{H^\pm_1} &=& 446\; [974],\;\;\;
M_{H^\pm_2} = 10433\; [10433].
%M_{H_1^{\pm \pm}}& =& 500,\;\;\;
%M_{H_2^{\pm \pm}}=500 \;\;\;
\label{B2MH2pp}
\end{eqnarray}  

This spectrum is obtained  with the following set of potential parameters ($v_R = 8\,\mathrm{TeV}$):
\begin{eqnarray}
\lambda_1 &=& 0.129\; [0.129],\;
\lambda_2 = 0\;[0],\\
\lambda_3 &=& 1\; [1]\;
\lambda_4 = 0\; [0], \\
\alpha_1 &=& 0\; [0], \;
\alpha_2 = 0\; [0], \;
\alpha_3 = 3.4\; [3.4], \\
\rho_1 &=& 5.7\; [5.7], \;
\rho_2 = 0.00115\; [0.00701], \\
\rho_3 &=& 11.405\; [11.428].
\label{potential_param_1}
\end{eqnarray}  

%\footnote{${p_{T_j}}$ cut is given as a regulator to make cross-section convergent and consistent. {\bf JD, reference? meaning? is it necessary to mention it?}}.
%At the LHC for $\sqrt{s}=14$, $33$, $100\,\mathrm{TeV}$ 
The  cross sections for the following process at the parton level with minimal imposed cuts  are given as follows:
\begin{eqnarray}\label{eq:xsection_1}
&&\sigma(p p \rightarrow H_{1/2}^{\pm \pm} H_{1/2}^{\mp \mp} j j \rightarrow \ell^{\pm} \ell^{\pm} \ell^{\mp} \ell^{\mp} jj)\\
&& =\left\{ \begin{array}{lcl}
4.04\;[0.12]\times 10^{-2}~ {\rm fb}     & {\rm for} & \sqrt{s}=14\;{\rm TeV}, \\
45.30\;[3.36]\times 10^{-2}~ {\rm fb}     & {\rm for} & \sqrt{s}=33\;{\rm TeV}, \\
282.80\;[31.76]\times 10^{-2}~ {\rm fb} & {\rm for} & \sqrt{s}=100\;{\rm TeV},
 \end{array} \right. \nonumber
\end{eqnarray}
where $\ell=e,\mu$. These minimal cuts are e.g. minimum $p_T$ cuts for leptons and jets such that they are identified as observable in the detector and do not contribute to missing energy. 

The result in Eq.~(\ref{eq:xsection_1}) is further processed using the VBF cuts   
\begin{eqnarray}\label{eq:xsection_1eff}
&&\sigma(p p \rightarrow H_{1/2}^{\pm \pm} H_{1/2}^{\mp \mp} j j \rightarrow \ell^{\pm} \ell^{\pm} \ell^{\mp} \ell^{\mp} jj)\\
&& =\left\{ \begin{array}{lcl}
0.54\;[0.01]\times 10^{-2}~ {\rm fb} & {\rm for} & \sqrt{s}=14\;{\rm TeV}, \\
6.21\;[0.40]\times 10^{-2}~ {\rm fb} & {\rm for} & \sqrt{s}=33\;{\rm TeV}, \\
37.01\;[3.54]\times 10^{-2}~ {\rm fb} & {\rm for} & \sqrt{s}=100\;{\rm TeV}.
 \end{array} \right. \nonumber
\end{eqnarray}

%\begin{eqnarray}\label{eq:xsection_1eff}
%&&\sigma(p p \rightarrow H_{1/2}^{\pm \pm} H_{1/2}^{\mp \mp} j j \rightarrow \ell^{\pm} %\ell^{\pm} \ell^{\mp} \ell^{\mp} jj)\nonumber\\
%&&= (?, ?, ?)\times 10^{-2}~ \rm{fb}. 
%\end{eqnarray}
For the sake of completeness let us display contributions from two intermediate channels\footnote{In \cite{Bambhaniya:2013wza} we wrongly assigned  $H_{1}^{\pm \pm}$ with right triplet and $H_{2}^{\pm \pm}$ with left triplet in Eq.~(A.13). 
%As a consequence, in that paper production of $H_{1}^{\pm \pm}$ for the considered processes (without jets)  has been mistakenly written as larger than a production of
%$H_{2}^{\pm \pm}$. 
We would like to thank Juan Carlos Vasquez for directing our attention to this fact as he considered this process in \cite{Vasquez:2014mxa}.}
%\begin{eqnarray}
%&&\sigma(p p \rightarrow H_{1}^{\pm \pm} H_{1}^{\mp \mp} j j)\nonumber\\ 
%&&=  (2.423,30.3,245.3) \times 10^{-2} ~\rm{fb},\label{eq:xsection_2a}\\
%&&\sigma(p p \rightarrow H_{2}^{\pm \pm} H_{2}^{\mp \mp} j j)\nonumber\\ 
%&&= (2.639,31.78,208.5) \times 10^{-2}~\rm{fb},\label{eq:xsection_2b}
%\end{eqnarray}
with the default cuts in Madgraph (MG):
\begin{eqnarray}\label{eq:xsection_1effH1}
&&\sigma(p p \rightarrow H_{1}^{\pm \pm} H_{1}^{\mp \mp} j j) \\
&& =\left\{ \begin{array}{lcl}
11.16\;[0.39]\times 10^{-2}~{\rm fb}   & {\rm for} & \sqrt{s}=14\;{\rm TeV}, \\
90.87\;[7.05]\times 10^{-2}~ {\rm fb} & {\rm for} & \sqrt{s}=33\;{\rm TeV}, \\
599.70\;[73.28]\times 10^{-2}~ {\rm fb}   & {\rm for} & \sqrt{s}=100\;{\rm TeV},
 \end{array} \right. \nonumber
\end{eqnarray}
and
\begin{eqnarray}\label{eq:xsection_1effH2}
&&\sigma(p p \rightarrow H_{2}^{\pm \pm} H_{2}^{\mp \mp} j j )\\
&& =\left\{ \begin{array}{lcl}
8.35\;[0.19]\times 10^{-2}~ {\rm fb} & {\rm for} & \sqrt{s}=14\;{\rm TeV}, \\
71.20\;[3.81]\times 10^{-2}~ {\rm fb}     & {\rm for} & \sqrt{s}=33\;{\rm TeV}, \\
401.40\;[37.43]\times 10^{-2}~ {\rm fb}     & {\rm for} & \sqrt{s}=100\;{\rm TeV}.
 \end{array} \right. \nonumber
\end{eqnarray}

%\begin{eqnarray}
%&&\sigma(p p \rightarrow H_{1}^{\pm \pm} H_{1}^{\mp \mp} j j)\nonumber\\ 
%&& = (11.16, 90.87, 599.7) \times 10^{-2} ~\rm{fb},\label{eq:xsection_2a_MG}\\
%&&\sigma(p p \rightarrow H_{2}^{\pm \pm} H_{2}^{\mp \mp} j j)\nonumber\\ 
%&& = (8.345, 71.2, 401.4) \times 10^{-2}~\rm{fb},\label{eq:xsection_2b_MG}
%\end{eqnarray}
%at the LHC with $\sqrt{s}=14$, $33$, $100$ TeV respectively.  
As one can see,  the cross sections in Eqs.~(\ref{eq:xsection_1effH1}) and (\ref{eq:xsection_1effH2}) 
are  larger than these given in Eq.~(\ref{eq:xsection_1}). The reason for this is while computing the cross section for the leptonic final state, i.e., Eq.~(\ref{eq:xsection_1eff}), all the selection cuts are incorporated and that reduces the cross section by a large amount.

Some technical details related to the computing method are in order here. At the MG level, one can control the gluon contributions using option \texttt{QCD=0}.  The cross section for $p p \rightarrow H_{1,2}^{\pm \pm} H_{1,2}^{\mp \mp} j j$ with switched off gluons turns out to be about $5$ times smaller than for that with gluons. 
%without gluons contributions. 
Hence, QCD contributions to that signal are really important. However, in both cases, distributions of the rapidity ($y$) of jets are quite different. Allowing for gluons, they are peaked around $y = 0$, which implies that jets are emitted mostly perpendicular to the beam. Otherwise, rapidities are peaked around $|y| \sim 3$ and $|y_1-y_2| \sim 5$;  i.e., there are two back-to-back jets emitted along the beam. Hence, setting \texttt{QCD=0} allows us to preselect processes which are consistent with VBF cuts. Effectively, it  shortens computing time\footnote{Typical run times for generating $5\cdot 10^4$ events of $p p \to H_{1/2}^{\pm \pm} H_{1/2}^{\mp \mp} j j$ and $p p \to  H_{1/2}^{\pm \pm} H_{1/2}^{\mp \mp} j j \rightarrow \ell^{\pm} \ell^{\pm} \ell^{\mp} \ell^{\mp} jj$  with \texttt{QCD=0} are, respectively, about $3\,$h and $54\,$h on $8$ core $3.4\,\mathrm{GHz}$  CPU.}.

%*********************************************************************************

Let us  comment on the $H^{\pm \pm}$  decay scenario used in the calculations. It is assumed that the decay of $H^{\pm \pm}$ is dominantly into a pair of the same sign and same-flavor charged leptons (for all possibilities within MLRSM, see \cite{Bambhaniya:2013wza}).  In other words, it is assumed that the Yukawa coupling matrix of doubly charged scalar $H_2^{\pm \pm}$ with charged leptons is diagonal.  Assuming no mixed leptonic decay modes ($e \mu$), i.e., no lepton flavor violation, the coupling of doubly charged scalar $H_2^{\pm \pm}$ with charged leptons in MLRSM is proportional to the heavy neutrino mass of the corresponding lepton generation. Thus the   $ee,\mu \mu$ decay modes will be larger compared to the $\tau \tau$ case if the first and second generations of right-handed neutrinos are more massive than the third-generation one. This point has been clarified and shown numerically in Fig.~2.5 in Ref.~\cite{Bambhaniya:2013wza}. In the present analysis, the masses of the first two generations of right-handed neutrinos are taken to be $3\,\mathrm{TeV}$ and the mass of the third one is at the level of $800\,\mathrm{GeV}$. As $v_R=8$ TeV, the Yukawa couplings are within the perturbative limit. If the $\tau \tau$ decay mode would be larger,  predictions given here should be rescaled properly using corresponding branching ratios (for instance, in the democratic three-generation case, the branching ratio for the $ee$ and $\mu \mu$ channels would be decreased by about $15 \%$ each). 

\begin{table}[h!]
\centering
\begin{tabular}{|c|c|c|c|}
\hline
Process: $ZZjj$ & Cross section [fb] & Cross section [fb] \\
 with $\sqrt{s}$  & at parton level       & after showering, \\
 in TeV               & in Madgraph          & and hadronization \\
 && in PYTHIA\\
\hline
14 & 0.115 & 0.003    \\
\hline
33 & 1.109 & 0.008    \\
\hline
100 & 4.794 & 0.038    \\
\hline
\end{tabular}
\caption{Standard Model cross section in fb for $\ell^{+} \ell^{-} \ell^{+} \ell^{-} j j$ final state and $\sqrt{s} = 14$, $33$, $100\,\mathrm{TeV}$ LHC. The cuts are suitably applied, see section \ref{sec:simulation_criteria}, to compute the SM background at the parton level and after incorporating showering and hadronization in PYTHIA. }
\label{table:lllljj_14TeV}
\end{table} 
%\end{document}

The SM background at the LHC for the signal $4\ell + 2$ jets  is accounted from the process $pp\rightarrow ZZ (\gamma \gamma, Z \gamma)jj\rightarrow \ell^{+} \ell^{-} \ell^{+} \ell^{-}jj$. We have noted that after implementation of the selection cuts, the dominant background is contributed from the $pp\rightarrow ZZjj\rightarrow \ell^{+} \ell^{-} \ell^{+} \ell^{-}jj$ process.

 For $\sqrt{s} = 14$, $33$, $100\,\mathrm{TeV}$ $pp$ collisions, the SM background is given in  Tab.~\ref{table:lllljj_14TeV} both at the parton level and after hadronization and passing through  implemented cuts.

We can see that the background is suppressed very effectively.
The results in Tab.~\ref{table:lllljj_14TeV} are obtained in the leading order;  electroweak corrections can change the results not more  than $10\%$ \cite{Catani:2009sm} which can change the significance of signals at the level of about one percent at most.

\begin{figure*}[htp]
%\begin{center}
%\includegraphics[width=5cm, angle=-90]{PLOTS/3000L/ssdl_lepton_distribution.eps}
%\includegraphics[width=5cm , angle=-90]{PLOTS/1000L/ssdl_lepton_distribution.eps} \vskip 0.1cm
%\includegraphics[height=7cm, width=7cm , angle=-90]{Significance_vs_Luminosity_LR_VBF_Mhcc500GeV.eps} 
\subfigure{\includegraphics[height=7.5cm, width=7cm , angle=-90]{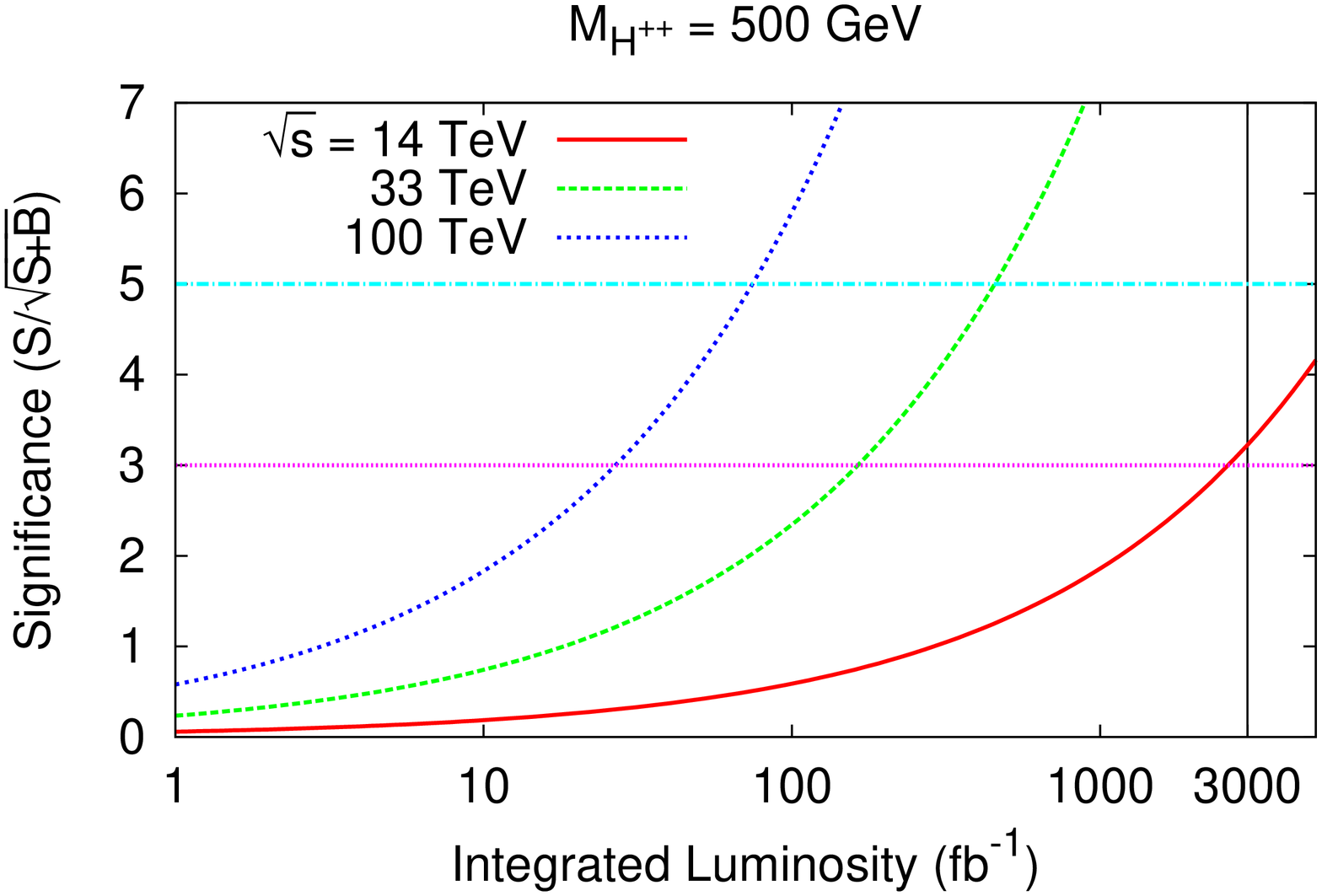}}\quad 
\subfigure{\includegraphics[height=7.5cm, width=7cm , angle=-90]{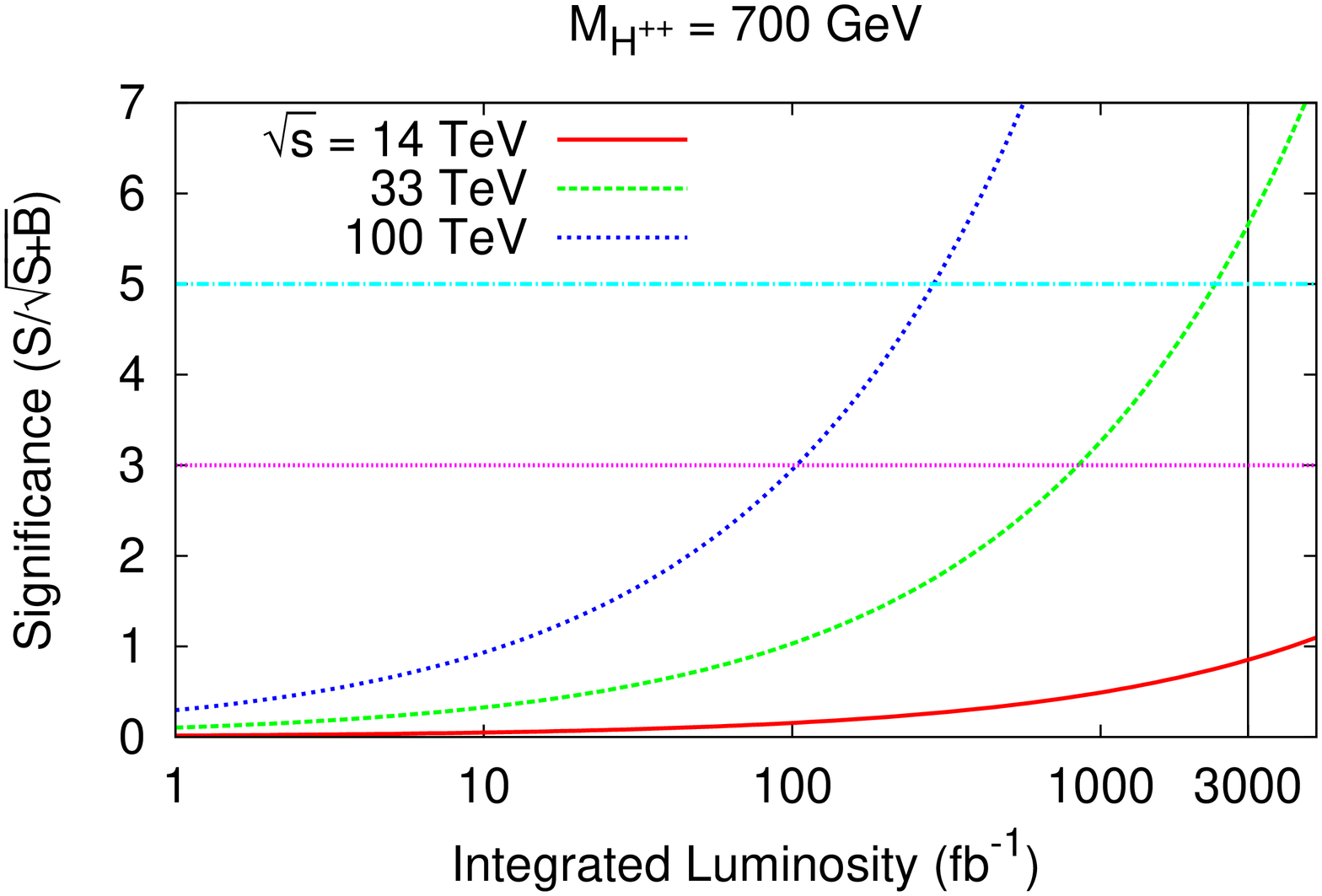}}\\
\subfigure{\includegraphics[height=7.5cm, width=7cm , angle=-90]{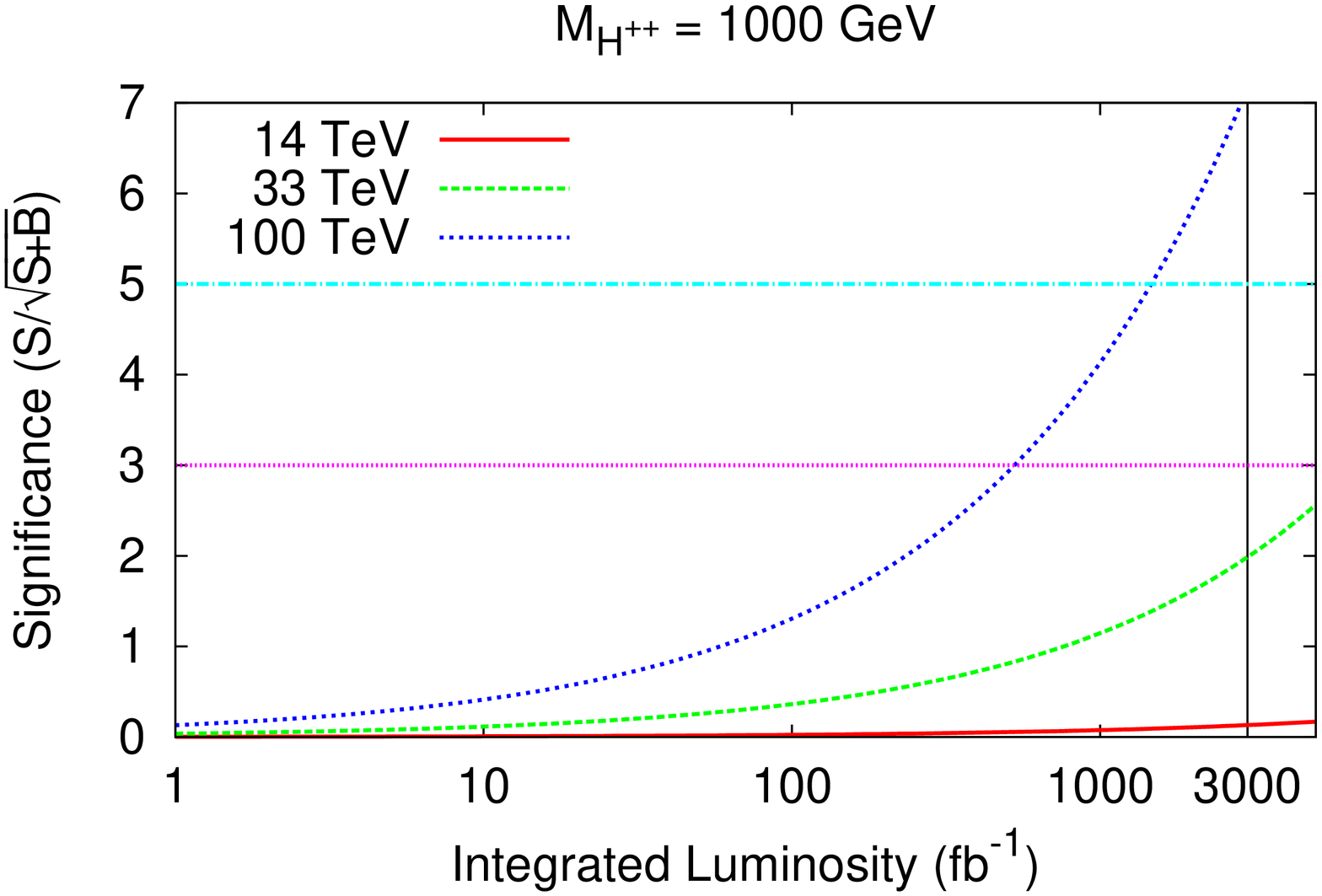}}\quad 
\subfigure{\includegraphics[height=7.5cm, width=7cm , angle=-90]{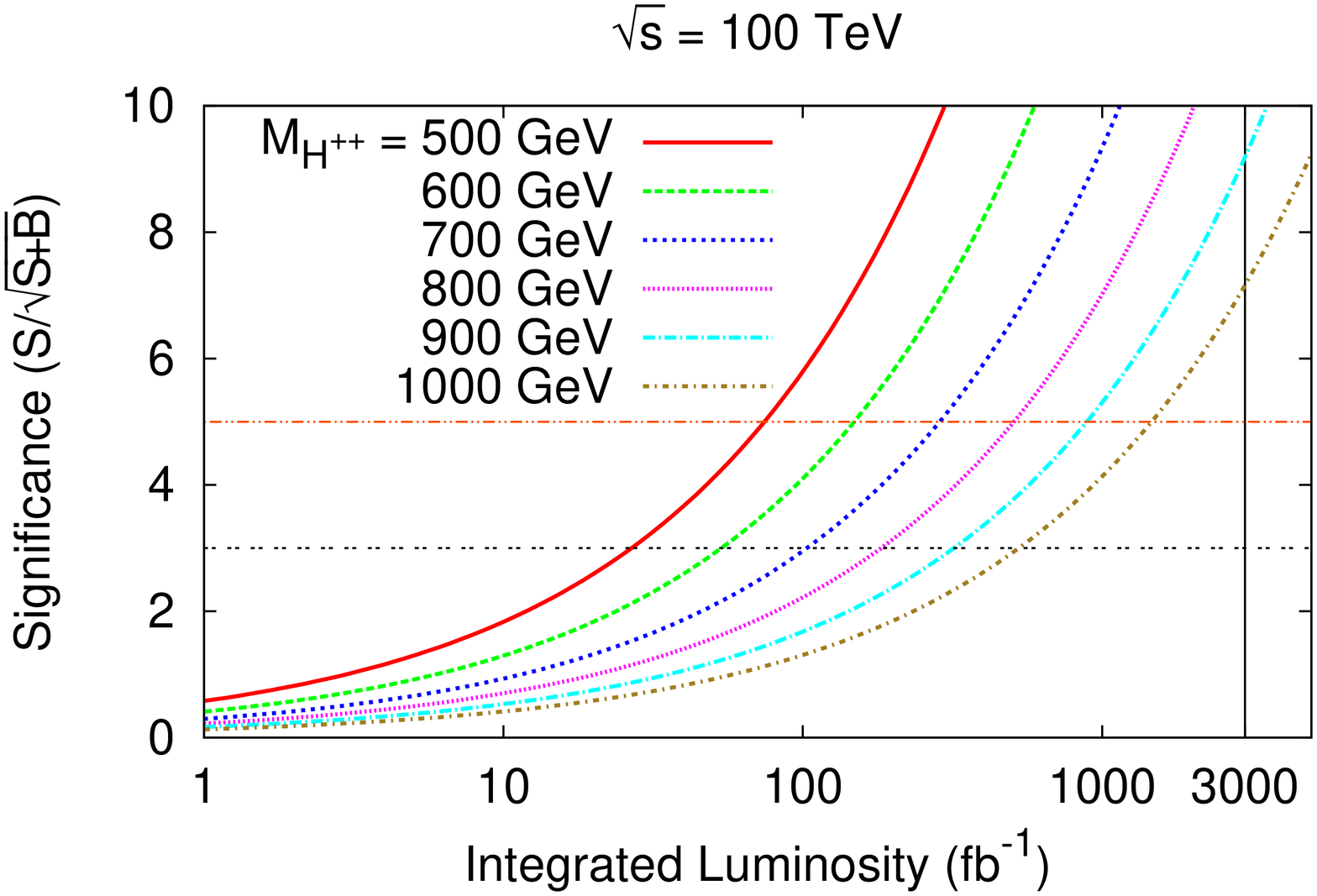}}
\caption{Variations of significance of the $signal$ with integrated luminosities for different energies of $pp$ colliders and various doubly charged Higgs boson masses.}
\label{fig:signif}
%\end{center}
\end{figure*}
%\begin{figure}[h!]
%\begin{center}
%\includegraphics[height=7.5cm, width=7cm , angle=-90]{Significance_vs_Luminosity_LR_VBF_Mhcc700GeV.eps}
%\end{center}
%\end{figure}
%\begin{figure}[h!]
%\begin{center}
%\includegraphics[height=7.5cm, width=7cm , angle=-90]{Significance_vs_Luminosity_LR_VBF_Mhcc1000GeV.eps}
%\end{center}
%\end{figure}
%\begin{figure}[h!]
%\begin{center}
%\includegraphics[height=7.5cm, width=7cm , angle=-90]{Significance_vs_Luminosity_LR_VBF_Mhcc800_900_1000GeV.eps}  
%\caption{Variations of significance of $signal$ with integrated luminosities for different energies of $pp$ colliders and various doubly charged Higgs boson masses.}
%\label{fig:signif}
%\end{center}
%\end{figure}

Finally, to judge  the strength of the MLRSM signals we decided to show the dependence of the significance of the result as a function 
of the integrated luminosity.
%in this channel with the luminosity. 
As can be seen in Fig.~\ref{fig:signif} (left-top),  a comfortable value of the significance at the level of $5$ can be reached for
 $M_{H^{\pm \pm}}=500\,\mathrm{GeV}$ in $pp$ collisions with 
\begin{itemize}
\item $\sqrt{s}=100$ TeV  and with 100 $\mathrm{fb}^{-1}$ integrated luminosity; 
\item   $\sqrt{s}=33$ TeV and with 700 $\mathrm{fb}^{-1}$ integrated luminosity.
\end{itemize}

No signal at this significance level ($\sim 5$) can be reached  with $\sqrt{s}=14\,\mathrm{TeV}$ $pp$ collisions, even if the integrated luminosity is around 3000 $\mathrm{fb}^{-1}$.

In Fig.~\ref{fig:signif} (right-top) we can see that doubly charged Higgs bosons with masses up to $M_{H^{\pm \pm}}=700\,\mathrm{GeV}$ with significance at the level of 5 can be probed for both center of mass energy 33 and 100 TeV with integrated luminosities around 3000 and 300 $\mathrm{fb}^{-1}$, respectively. 
In  Fig.~\ref{fig:signif} (left-bottom) it is evident that the 1 TeV doubly charged scalar can be probed with a significance of 5 only with 100 the TeV collider with luminosity at least 1000 $\mathrm{fb}^{-1}$.
The Fig.~\ref{fig:signif} (right-bottom)  summarizes the situation for the FCC-hh collider option for three different sets of masses of doubly charged scalars: 500, 600, 700, 800, 900 and 1000 GeV. This figure also
shows that  significance at the level of 7 can be reached for $M_{H^{\pm \pm}}=1\,\mathrm{TeV}$ and $\sqrt{s}=100$ TeV with integrated luminosities around 3000 $\mathrm{fb}^{-1}$.   We can see that this collider opens up a very wide range of Higgs boson masses which can be explored.

\section{Conclusions and Outlook}

In this paper we have considered production and decays of a  pair of doubly charged Higgs bosons through vector boson fusion within the MLRSM framework. 
To do so we have evaluated suitable benchmark points for masses of Higgs bosons, which are in agreement with  several constraints coming from FCNC, vacuum stability, LEPII and recent ATLAS searches on doubly charged scalars. There are strong relations among masses of doubly and singly charged and neutral scalars which prevent us from choosing their individual values freely, leaving us with the suitable  benchmarks we are using in this paper. We have further noted and shown  that the  splitting between the doubly ($H_1^{\pm \pm}$ ) and singly ($H_1^\pm$) charged scalars  is less than $M_{W_1}$, irrespective of the $SU(2)_R$ breaking scale. Thus, the on-shell decay $H_1^{\pm \pm} \to H_1^{\pm } W_1^{\pm}$ is protected and  the decay branching ratio of the doubly charged scalar $H_1^{\pm \pm}$ is affected. 

After settling these issues regarding the spectrum, we have computed the signal cross section for the process $p p \rightarrow H_{1/2}^{\pm \pm} H_{1/2}^{\mp \mp} j j \rightarrow \ell^{\pm} \ell^{\pm} \ell^{\mp} \ell^{\mp} jj$ using realistic cuts. The necessary SM background for this final state is also evaluated. It has been shown that LHC2 even with high integrated luminosity will be not be sensitive to the  VBF-like signals $H_{1/2}^{\pm \pm} H_{1/2}^{\mp \mp} j j$, even with relatively light doubly charged Higgs bosons (say, $\sim 500\,\mathrm{GeV}$).  We have shown that much better perspective exists for the future FCC colliders with center of mass energies 33 and (or) $100\,\mathrm{TeV}$. 

 In passing we would like to mention that we have used the VBF cuts as adopted in \cite{Dutta:2014dba}. We have compared the ATLAS and CMS (tight and loose) suggested cuts which are not very widely different from significance point of view. 
 %For further improved predictions, in future the analysis including NLO corrections should be undertaken, for instance it is known that electroweak corrections to VBF signals are within 10\% level \cite{Catani:2009sm} and might have a mild impact on  significance computed at LO.

Let us conclude with a comparative comment:  MLRSM VBF-like signals connected with $H_2^{\pm \pm}$ scalar production 
(which is part of the right-handed triplet) is comparable with the  $H_1^{\pm \pm}$ scalar production, see Eqs.~(\ref{eq:xsection_1effH1}) and (\ref{eq:xsection_1effH2}). Thus the cross section for signal events are larger compare to that for Type-II seesaw scenario with same masses for triplet scalars.
It  may give a chance to disentangle between MLRSM and SM with an additional triplet, e.g. the Higgs Triplet Model \cite{Melfo:2011nx, Han:2007bk, Perez:2008ha, Englert:2013zpa, kang:2014jia, Kanemura:2014goa}, though detailed analyses are needed for comparison. 

\section*{Acknowledgements}
The authors would like to thank Benjamin Fuks for sharing with them a modified version 
%2.0.31 
of FeynRules. 
This work is  supported  by the Department of Science and Technology, Government of India under the Grant Agreement No. IFA12-PH-34 (INSPIRE Faculty Award) and the Polish National Science Centre (NCN) under Grant Agreement No. DEC-2013/11/B/ST2/04023 and under Postdoctoral Grant No. DEC-2012/04/S/ST2/00003. The work of R.S.  is supported by Science and Engineering Research Canada (NSERC). 

\appendix 
%\providecommand{\href}[2]{#2}
%\addcontentsline{toc}{section}{References}
%\bibliographystyle{JHEP}
%\bibliography{LRref}

\end{document}